%% file: usenix2021.tex
\documentclass[letterpaper,twocolumn,10pt]{article}
\usepackage{usenix2019_v3_hotos}

\usepackage{tikz}
\usepackage{amsmath}

\usepackage{booktabs}

\usepackage[compact]{titlesec}
\titlespacing*{\section}{0pt}{6pt plus 4pt minus 2pt}{2pt plus 2pt minus 2pt}
\titlespacing*{\subsection}{0pt}{4pt plus 2pt minus 1pt}{2pt plus 1pt minus 1pt}
\titlespacing*{\subsubsection}{0pt}{4pt plus 2pt minus 1pt}{2pt plus 1pt minus 1pt}

\begin{document}

%don't want date printed
\date{}

\title{{\bf SLO beyond the Hardware Isolation Limits}}

%for single author (just remove % characters)
\author{
{\rm Haoran Qiu}\\
UIUC
\and
{\rm Yongzhou Chen}\\
UIUC
\and
{\rm Tianyin Xu}\\
UIUC
\and
{\rm Zbigniew T. Kalbarczyk}\\
UIUC
\and
{\rm Ravishankar K. Iyer}\\
UIUC
} % end author

\maketitle

\input{000-abstract}
\input{001-intro}
\input{002-related}
\input{003-overview}
\input{004-eval}
\input{005-discussion}
\bibliographystyle{plain}
\bibliography{reference}

\end{document}

%% file: 000-abstract.tex
\begin{abstract}
% Performance unpredictability is a "deadly sin" for real-time operation and most user-facing interactive cloud services.
Performance isolation is a keystone for SLO guarantees with shared resources in cloud and datacenter environments.
To meet SLO requirements, the state of the art relies on hardware QoS support (e.g., Intel RDT) to allocate shared resources such as last-level caches and memory bandwidth for co-located latency-critical applications.
As a result, the number of latency-critical applications that can be deployed on a physical machine is bounded by the hardware allocation capability.
Unfortunately, such hardware capability is very limited.
For example, Intel Xeon E5 v3 processors support at most four partitions for last-level caches, i.e., at most four applications can have dedicated resource allocation.
This paper discusses the feasibility and unexplored challenges of providing SLO guarantees beyond the limits of hardware capability.
We present CoCo to show the feasibility and the benefits.
CoCo schedules applications to time-share interference-free partitions as a transparent software layer.
Our evaluation shows that CoCo outperforms non-partitioned and the round-robin approaches by up to 9$\times$ and 1.2$\times$.

% (tianyin) This argument needs to be backed up by discusssions
% However, with the increasing popularity of microservices and Function-as-a-Service paradigm, the number of containers consolidated together increases significantly.
\end{abstract}

%% file: 001-intro.tex
\section{Introduction}
\label{sec:introduction}

% 1. SLO guarantee is achieved by reservation/isolation. The idea is that the target workload would not be interfered by noisy neighbors, thus achieving predictable performance.

% As an increasing number and variety of enterprises are moving workloads to cloud platforms, the runtime behavior of an application is affected by other applications running concurrently and thus, competing for shared resources (e.g, CPU cycles, cache, main memory).
% With dynamic interference, it is rather difficult to predict the completion time~\cite{parties,fried2020caladan,clite,novakovic2013deepdive,qiu2020firm}, which is crucial for real-time operations and most user-facing interactive cloud services with strict SLOs~\cite{loss}.
% Therefore, exhibiting a predictable performance requires an application to be isolated from others it shares resources with.

Service-level objectives (SLOs) guarantee in shared environments such as cloud and datacenters is predominantly done by resource isolation~\cite{heracles,parties,clite,copart,qiu2020firm,javadi2019scavenger,tootoonchian2018resq,zhu2016dirigent,herdrich2016cache}.
Without resource isolation, the performance of an application can be interfered with and impaired by co-located applications that compete for shared resources such as CPU, last-level caches, and memory bandwidth.
In shared cloud and datacenter environments, such interference is inevitable and hard to mitigate, leading to unpredictable performance~\cite{parties,clite,novakovic2013deepdive,qiu2020firm,blagodurov2015multi,zhuravlev2012survey,maricq2018taming}.

% 2. This is achieved by hardware or even physical isolation. Google and FB choose to run LC jobs with dedicated machines (FB even chooses wimpy machines for their infra, because of this). This results in significant resource waste in modern data centers.
% 3. Modern architecture features such as CAT and MBA enable fine-grained resource isolation capabilities based on the CLOS allocation primitive.
% 4. There have been research efforts leveraging CAT/MBA for fine-grained allocation with SLO guarantee (cite and praise them).

% To meet the performance isolation condition without overprovisioning, several hardware mechanisms and kernel features that partition resources are developed.
% For example, cgroups~\cite{cgroups} enables isolation of CPU cores (and other resource types) to containers.
% Intel Cache Allocation Technology (CAT)~\cite{herdrich2016cache} and Memory Bandwidth Allocation (MBA)~\cite{mba} provide LLC capacity partitioning and memory bandwidth throttling to multiple isolated or overlapping Classes of Service (CLOS).
% Relying on resource partitioning, SLO-aware resource management proposed in previous work~\cite{heracles,parties,clite,copart,qiu2020firm,javadi2019scavenger,tootoonchian2018resq,zhu2016dirigent} is able to guarantee strict SLO requirements of latency-critical (LC) services while maximizing resources allocated to best-effort (BE) tasks.

To provide performance isolation, hardware QoS support~\cite{hardwareqos,amd} and OS kernel extensions are developed.
The key idea is to reserve and allocate dedicated resources for applications with strict SLO requirements, i.e., latency-critical (LC) applications.
For example, Intel RDT (Resource Director Technologies) provides cores with fine-grained isolation support for last-level caches~\cite{herdrich2016cache} and memory bandwidth~\cite{mba}.
Control groups (cgroups)~\cite{cgroups,blagodurov2013maximizing} can be used to leverage the hardware support by assigning tasks to cores, and to limit the I/O bandwidth and memory capacity that are available to a group of tasks.
% To exploit the hardware support, cgroups~\cite{cgroups} is developed to assign tasks to run on isolated cores, which also controls memory usage and throttle resources such as I/O at the OS level.
% [DONE] \tianyin{There needs a transition from cgroups to hardware features.}
The state-of-the-art resource management solutions~\cite{heracles,parties,clite,copart,qiu2020firm,javadi2019scavenger,zhu2016dirigent} all rely on those supports to meet SLO requirements by reserving dedicated resources for LC applications; meanwhile, the unreserved resources could be shared among best-effort applications (e.g., batch jobs) to improve resource utilization.

% 5. However, such SLO guarantee is fundamentally limited by the hardware isolation support (e.g., at most 3 LC jobs can be run at the same time no matter how many cores/memory you have).
% 6. The current hardware support cannot accommodate the needs of modern computing paradigms (e.g., deployment scale required by serverless and FaaS).
%However, the number of LC services supported is fundamentally limited by the hardware isolation support.
% For example, the number of LC services managed in PARTIES~\cite{parties} cannot exceed the maximum number of CLOS (which is four even in the Intel Xeon E5 family processors) supported in Intel CAT/MBA because each LC service is assigned to one CLOS.
% Recent transition to microservices and Function-as-a-Service (FaaS) paradigms significantly increases the number of containers deployed in host machines.
% The current hardware support cannot accommodate this trends of deployment scaling no matter how many cores or memory each machine has.

Currently, the number of LC applications that can be deployed on a physical machine is bounded by the hardware allocation capability.
For example, PARTIES~\cite{parties}, the state-of-the-art resource partitioning technique, cannot support more applications than the number of hardware partitions (defined as CLOSs, Classes of Services, in Intel RDT).
Unfortunately, the hardware allocation capability is very limited.
For example, Intel Xeon E5 v3 family processors only support four CLOSs~\cite{rdtnumclos}.
That means that at most four LC applications can be deployed no matter the number of CPU cores or the memory size.

The fact that resource isolation is bounded by the limited hardware allocation capability hinders improvements of resource utilization.
% On the other hand, it can lead to lower resource utilization efficiency.
For instance, four LC applications are consolidated on a physical machine with a 4-CLOS processor. Each application is assigned 10\% of the last-level cache (in total 40\%) to guarantee its SLO.
% running 4 LC applications running on a 4-CLOS processor, each of the application is assigned 10\% of last-level cache capacity (in total 40\%) to guarantee SLOs.
The wasted 60\% of the cache capacity could otherwise be utilized to support more LC applications.
The problem becomes emergent, considering recent trends of microservices and serverless computing which encourages ``micro'' applications with stringent SLO requirements~\cite{gan2018architectural,shahrad2019architectural,yu2020characterizing}.
Certainly, a fundamental solution to this problem is to redesign hardware QoS support to expose as many CLOSs as possible. However, deploying new hardware takes time, let alone the fundamental challenges in scaling cache associativity (discussed in \S\ref{sec:conclusion}).
We believe that the discrepancy between the hardware QoS support and the scale of CPU cores will continue to exist if not become worse.

% 7. We advocate that SLO guarantee should not be limited by the hardware limits, especially given that existing hardware features (e.g., CAT/MBA) does not provide sufficient capability.

This paper investigates the feasibility of providing SLO guarantees beyond the limits of hardware allocation capability, and discusses the related issues and limitations (see \S\ref{sec:discussion}).
Our aim is to support SLO guarantees for more applications than the number of CLOSs provided by the hardware and to exploit unused hardware resources.
We believe that providing SLO guarantees beyond the hardware isolation limits is significant---instead of passively waiting for hardware innovations, we seek for software-only solutions to efficiently utilize the CLOSs available on the physical machine.
%, given that existing hardware features (e.g., Intel RDT) do not offer desired capabilities. 

% We advocate that SLO guarantees should not be limited by the hardware boundary, especially given that existing hardware features (e.g., CAT/MBA) do not provide sufficient capability.
% Instead of slowing down the software evolution (e.g., microservices and serverless computing) and waiting for hardware upgrades to catch up, can we provide SLO guarantees to tens of containers with only limited number of resource partitions on one physical machine?

% 8. We explore the feasibility and the benefit of providing SLO guarantee beyond the hardware limits (four CLOS). // Seriously, we’d better not wait for hardware upgrades (and hardware support can never catch up with software ambition).
% 9. We demonstrate the feasibility and the benefits by designing CoCo... your amazing solution

Our key insight is to treat CLOS-based partitions as scarce resources and enable {\it time-sharing} among co-located applications without breaking application SLOs.
This is in contrast to state-of-the-art approaches (e.g., PARTIES~\cite{parties}) that statically assign each application to a partition and increase or decrease the amount of resources allocated to that partition so as to guarantee the application SLO.
% The shareability depends on the characteristics of consolidated cloud services and limits the upper-bound of the affordable client load for cloud services without SLO violation.
% This project aims to bridge the gap of scheduling multiple co-located LC services beyond the limited number of supported CLOS to optimize for performance and resource utilization.
Since each application has different sensitivities with regards to the amount of resources and time slices associated with a partition,
the key challenge is to design the time-shareability based on the characteristics of co-located applications, and not to trade off SLOs for improving resource utilization.
% (measured as its capacity of serving client requests)
% affordable client load upper-bound for each application on a physical machine is affected by time-sharing.
% Time-sharing a partition cannot trade SLO violations for improving resource utilization.
% Although the affordable client load on a single machine is not able to fulfill the total client traffic, cloud on-demand scaling and load-balancing can easily offset the cost.
% [haoran] The affordable client load is not infinity, there's a point at which the SLO can no longer be maintained. Similarly in our case, there's a point at which the application cannot time-share its partitions, otherwise the SLO will be violated, and we call it the upper bound
% [haoran] We increase the utilization on one physical machine, but the affordable client load is affected, so the load needs to be balanced on multiple such physical machines to fulfill client needs. E.g., for a webserver, the real client load is 100 req/s, one machine #1, the upper bound is 40 req/s (without breaking the SLO of this 40% requests), machine #2 is 30 req/s, and machine #3 is 30req/s. In total they fulfill the 100req/s.

% Moreover, the application sensitivity can be affected by hidden system resources such as PCIe bandwidth contention.

% This project aims to bridge the gap of scheduling multiple co-located LC services beyond the limited number of supported CLOS to optimize for performance and resource utilization.

We present CoCo, 
% as the first step in the direction to realize the feasibility and demonstrate the benefit. CoCo is 
a coordinated container scheduler that transparently allows tens of SLO-oriented LC workloads to share a limited number of partitions of last-level caches and memory bandwidth.
%In the offline phase,
CoCo profiles the sensitivity of the target workload regarding resource limits.
% assign workloads with the least interference in one CLOS during their runtime.
% In the online phase,
CoCo configures the CLOS partitions based on the sensitivity characteristics and dynamically schedules workloads to different partitions based on their SLO requirements.
The core of CoCo is a novel application-aware time-sharing algorithm that maximizes the overall performance and resource utilization (see \S\ref{sec:design}).
We implemented a prototype of CoCo as a user-level runtime system for Linux, which requires no modification to the underlying kernel or co-located applications.\footnote{CoCo is open-sourced and available at: \url{https://anonymous.4open.science/r/a3f84af9-5901-4400-89dd-98b3ef15cc1d/}.}
Our evaluation shows that CoCo improves the maximum affordable client load without SLO violations by up to 9$\times$ and 1.2$\times$, compared with non-partitioned and the round-robin approach~\cite{tavakkol2018flin}.
CoCo is the first step in the direction of providing SLOs beyond the hardware limits but there are still unexplored challenges that need to be addressed (see \S\ref{sec:discussion}).

%% file: 002-related.tex
\section{Hardware QoS Support}
\label{sec:related_work}

% \subsection{Intel CAT and MBA}
% \textbf{Intel CAT and MBA.}

As hardware manufacturers continue to add additional cores onto processors, more workloads can be consolidated together.
However, co-located workloads could contend for shared resources, such as last-level caches (LLC) and memory-bandwidth.
For LC workloads, the unexpected delay for fetching data from the main memory (due to cache contention and invalidation) can negatively impact performance.
Hardware QoS support is recently provided by many hardware manufacturers (e.g, Intel~\cite{herdrich2016cache,mba}, AMD~\cite{amd}, ARM~\cite{wang2017swap}, IBM~\cite{ibm}) to support resource isolation.
Most existing hardware QoS support is for LLC cache allocation enforcement, LLC cache occupancy monitoring, LLC code-data prioritization, and memory bandwidth enforcement/allocation.
% Intel has developed Cache Allocation Technology (CAT) to enable more control over the LLC cache and how cores allocate into it.
% Using CAT, the system administrator can reserve portions of the cache for individual cores so that only these cores can allocate into them. As a result, other applications may not evict cache lines from these reserved portions of the cache via general use of the caches.

% \tianyin{@Haoran: you may want to write in a more generic way, rather than giving the feeling that your idea is specific to CAT/MBA. Two questions in my mind:
% (1) Does ARM has equivalent features of CLOS?
% Searched a bit, ARM processors have way-partitioning feature for L2 cache, AMD processors do have this CLOS-based partitioning of caches and memory bandwidth
% (2) Is there anything more than CAT/MBA on Intel?}
% (a) There's also Code and Data Prioritization (CDP). It's further partitioning data and code, allocating them isolated placement in the caches.
% (b) And pseudo-locking, it essentially locks (softly) the data of a core in LLC https://github.com/intel/intel-cmt-cat/tree/master/examples/c/PSEUDO_LOCK
% (c) CMT and MBM are cache/memory monitoring features provided by Intel (hardware counters)
% I think they (a, b, c) are not too related to our topic.
% (d) CAT can also be used to partition L2 cache
% But for resource partitioning, CAT and MBA are the only two features

% AMD enables cache and memory bandwidth partitioning up and down its Zen 2 processor family
For instance, Intel's CAT~\cite{herdrich2016cache} and MBA~\cite{mba} are developed for addressing the shared resource contention for LLC and main memory bandwidth respectively.
The goal of CAT/MBA is to enable resource allocation based on partitions or CLOSs.
Similar CLOS-based resource partitioning features are also provided by AMD processors~\cite{amd}.
The processor exposes a set of CLOSs into which applications (or individual threads) can be assigned.
A given CLOS used for CAT means the same thing as a CLOS used for
MBA.
The number of CLOSs are typically limited by the hardware.
For example, both Intel(R) Xeon(R) E5 v3 family processors and AMD EPYC 7002 series only support at most 4 CLOSs~\cite{rdtnumclos}.

For CAT, a capacity bit-mask provides a hint to the hardware indicating the cache space an application should be limited to (not necessarily way partitioning).
A mask bit set to ``1'' specifies that a particular CLOS can access the cache subset represented by that bit and vice versa.
For MBA, a programmable request rate controller is introduced between the cores and the high-speed interconnect to control memory accesses.
It enables indirect control over memory bandwidth for cores over-utilizing bandwidth relative to their CLOS configuration.
The memory bandwidth throttling value is of the format of a percentage (with 100\% meaning full access) and the limit and granularity of MBA are machine-related.

%% file: 003-overview.tex
\section{CoCo: SLO-aware Partition Sharing}
\label{sec:design}

We realize our idea of sharing CLOS partitions in CoCo, a coordinated container scheduler that transparently enables running tens of SLO-oriented LC workloads beyond the hardware isolation limits.
CoCo makes interference-free CLOS partitions shareable in a way that no SLO of co-located workloads would be violated, without overprovisioning.
Instead of statically assigning an LC workload to a partition and controlling the amount of resources allocated to the partition~\cite{parties,heracles,clite,copart}, CoCo treats each partition as a scarce resource and enables time-sharing the resource among co-located workloads.
% It dynamically configures the system state (i.e., through simultaneous LLC and memory bandwidth partitioning provided by Intel CAT and MBA)
% CoCo is a coordinated technique which simultaneously partitions the LLC and memory bandwidth across the consolidated applications.

CoCo is application-aware.
It characterizes the sensitivity of workloads with regard to resource limits through profiling (\S\ref{sec:overview-profiling}).
CoCo does this by running the workload along with resource limiters which throttling resource access to simulate performance interference.
% CoCo also dynamically analyzes the performance characteristics of each co-located application with the allocated resources.
% Figure \ref{fig:overview} shows the workflow diagram of CoCo.
CoCo will dynamically configure the CLOS resource allocation with Intel CAT/MBA and schedule containers to share the time-slices of each CLOS without violating SLOs in the runtime.
This is done based on the profiled sensitivity characteristics (\S\ref{sec:overview-algorithm}).
Finally, the scheduler in CoCo adjusts to workload changes by adjusting parameters based on runtime monitoring.

\subsection{Profiling Sensitivity}
\label{sec:overview-profiling}

% talk about motivation for this offline profiling (i.e., sensitivity varies with application workloads and resource types; put the six figures here)
% To assign the proper hardware resource to all those LC applications without overprovisioning, it is necessary to get the application characteristics with offline profiling.
The objective of profiling is to understand the sensitivity of the target workload to the scarcity of the same type of resource~\cite{parties,heracles}.
% \tianyin{is profiling needed by parities/heracles?}
% In addition, different applications can have different sensitivities 
For example, with highest contention intensity in terms of last-level cache (LLC), the 99\% tail latency is 8$\times$ of the median latency in Memcached, while the 99\% tail latency is 28$\times$ of the median latency in Nginx.
Intuitively, the scheduler should allocate larger time slices to Nginx workloads to be associated with the CLOS with more LLC ways, compared to Memcached workloads.
One workload could have different sensitivities to different resource shortages as well~\cite{heracles,parties}.
For Nginx, the performance degrades the same facing both memory bandwidth contention or LLC contention;
but LLC shortage is more disastrous compared with memory bandwidth shortage for Memcached and MongoDB workloads.
With this sensitivity profiling, the scheduler can optimize for both performance and resource utilization efficiency.

% talk about how we do offline profiling and how we use this results as weights to assist online dynamic partitioning
To quantify the subtle, diverse sensitivities discussed above, it requires a metric that can be measured simply and practically. CoCo uses a metric termed \textit{slowdown}.
Assume that $N$ workloads are co-located on the same physical server and that the CPU equipped in the server provides the LLC with $L$ ways, the maximum memory bandwidth of $B$, and the maximum MBA level of 100\% (i.e., no throttling).
The resource allocation state \(S_i\) of workload \(i\) (\(i \in [0, N - 1]\)) is defined as \((l_i, m_i)\), where \(l_i\) and \(m_i\) denote the LLC ways and MBA levels allocated to the workload \(i\).
The slowdown of workload \(i\) with the resource allocation state \(S_i\) is defined in Eq. \ref{eq:slowdown}, where \(SL_{full}\) and \(SL_{S_i}\) denote the load which sustains the SLO when the workload is allocated full resources and the resources specified in state \(S_i\), respectively.
\vspace{-5pt}
\begin{equation}
\label{eq:slowdown}
    slowdown_{S_i} = \frac{SL_{full}}{SL_{S_i}}
    \vspace{-5pt}
\end{equation}
Given the same system state $S_i$, the higher the slowdown value is, the more sensitive that the workload $i$ is when facing the LLC or memory bandwidth throttling.
Consequently, a workload with a higher slowdown value should be allocated to a partition with larger capacity and longer time window, compared with other co-located workloads.

The workloads are profiled with a resource limiter starting from ($l_i = L$, $m_i = 100\%$) and reducing the allocated LLC and memory bandwidth partition step by step (step size is the minimum adjustment configuration defined by CAT/MBA on each machine).
For each step, the slowdown values of the workload are calculated. 

\subsection{Dynamic Partitioning}
\label{sec:overview-algorithm}

Dynamic partitioning configures container-to-core and core-to-CLOS mappings to meet SLOs of all co-located workloads.
% talk about the algorithm MQ-WRR
The scheduling algorithm designed and implemented in CoCo is a multi-level queue-based weighted round-robin (MQ-WRR) where the design requirements are:
(i) to give preference to workloads with high slowdown values,
(ii) to take the difference of workload sensitivity to shared-resources into account, and
(iii) to avoid starvation.
The CLOS configurations in Intel CAT/MBA are modeled as queues.
Each CLOS is associated with a subset of caches that do not overlapp with another CLOS.
There is one CLOS reserved particularly for the scheduling process of CoCo and other background jobs (e.g., system jobs) running on the physical server that are not latency-critical.
All other CLOSs are for scheduling of co-located LC workloads running in containers.

\begin{figure}
    \centering
    \includegraphics[width=0.99\linewidth]{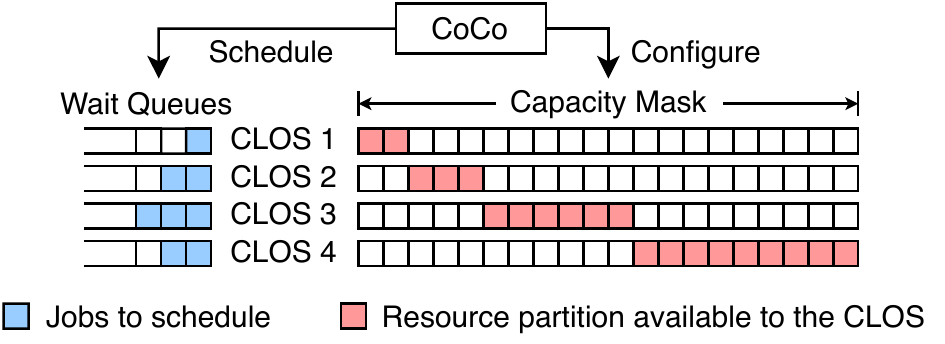}
    \caption{An example of CoCo scheduling.}
    \label{fig:scheduling}
\end{figure}

Figure~\ref{fig:scheduling} gives an example of CoCo scheduling using Intel CAT.
Each CLOS is configured using capacity bit-masks which represent cache capacity and indicate the degree of isolation between classes.
Once CAT is configured, the processor allows access to portions of the cache according to the established CLOS.
We use \texttt{CLOS1} for running background jobs that do not have critical SLOs, as well as CoCo’s scheduling process.
\texttt{CLOS2--4} are configured to have 3, 6, and 9 bits respectively.
Each CLOS has a wait queue consisting of containers that are going to be scheduled to the CLOS in the next time slice.
Each CLOS also has a working set (which is usually singleton for minimizing interference; but it can also serve two containers that are sensitive to different types of shared-resources) consisting of the container that is running right now.
CoCo is scheduling the containers in the weighted round-robin manner onto the multi-level queues.
Among $N$ workloads, the scheduling time slice on a CLOS (with state $s$) associated to workload $i$ is set to be $weight_i$, which is defined as:
\begin{equation}
    weight_i = \frac{slowdown_{S_i}}{\sum_{k=0}^N slowdown_{S_k}}
\end{equation}

Memory bandwidth partitioning using MBA is similar to cache partitioning.
Instead of setting bit-masks, MBA allows directly indicating the percentages of allowed memory bandwidth (an integer between 0 and 100).

If no schedule can meet all SLOs, admission control will kick in to keep only affordable workloads.

\subsection{Handling Conflicts}

% \textbf{Conflicts between Intel CAT and MBA.}
Although Intel CAT and MBA provide trustworthy partitioning of LLC and memory bandwidth to each CLOS, there may be conflicts when configuring CAT and MBA at the same time.
Since MBA uses a programmable rate controller between the cores and the interconnect, LLC and memory controller, bandwidth to LLC may also be reduced when reducing memory bandwidth using MBA.
For instance, the effect of increasing the bit length of the capacity bit-mask for \texttt{CLOS1} has a higher chance to be canceled out by setting a lower throttling value for \texttt{CLOS1}.
Therefore, it is hard to throttle memory-bandwidth-intense workloads which also uses the off-core caches effectively.

% However, coordination should be made between CAT and MBA to avoid conflicts (recall Section \ref{sec:challenges}).
To avoid partition conflicts between CAT and MBA, CoCo always increases or decreases memory bandwidth together with LLC capacity.
Otherwise, throttling bandwidth-intensive workloads, which also use the off-core caches, will be compromised by increasing memory bandwidth but decreasing LLC capacity.
As shown in \S\ref{sec:eval-e2e}, monotonically configuring CAT and MBA
yields a 29\% improvement of affordable client load.

% \textbf{Cache Misses After CLOS Migration.}
Once a set of CLOSs are configured by setting the bit-masks, the hardware will get the hint on how to partition the cache space for each CLOS.
After a change of CLOS configuration, the changed cache space ($\Delta C$ in terms of cache ways) will be added to or removed from the CLOS.
However, the cached pages in the cache space $\Delta C$ will not be flushed, which limits the new owner of the CLOS (which $\Delta C$ belongs to) from using the newly allocated cache space.
Similarly, a change of CLOS-process association has the same problem.
For instance, \texttt{CLOS-A} and \texttt{CLOS-B} own cache partition $C_a$ and $C_b$ respectively.
After a configuration change, \texttt{CLOS-A} owns $C_b$ and \texttt{CLOS-B} owns $C_a$.
But cache hits from \texttt{CLOS-A} can still lead to access to $C_a$, which is no longer owned by \texttt{CLOS-A} and thus limits the usage of $C_a$ by \texttt{CLOS-B}. % (leading to more cache misses).
If the access frequency from \texttt{CLOS-A} is high, the least-frequently-used cache replacement policy does not help with CLOS migration.

To resolve cache misses after CLOS migration, CoCo flushes and invalidates the cache in the original cache partition for each CLOS after each configuration change by paying up to 6.1\% overhead of cache warmup 
 (\S\ref{sec:eval-e2e}).

% Lastly, CoCo also takes NUMA nodes into consideration by consolidating threads of one application on the same NUMA nodes and never migrate threads across different NUMA nodes.
% In addition, cache partition in CoCo does not cross different NUMA nodes either.

\subsection{Discussion}
\label{sec:discussion}

CoCo is the first step in the direction towards time-sharing of CLOSs in hardware QoS support without violating SLOs, thus improving resource utilization on physical machines.
There are still unexplored challenges that need to be addressed.
First, just like statically allocating an LC workload to a CLOS, it cannot serve unlimited client load without SLO violations. We call the point where the client load is no longer affordable the \textit{upper bound}.
The application is usually scaled out to more machines when the client load is beyond the upper bound. 
By time-sharing, the upper bound for each LC workload may be reduced;
but we believe the cost can be offset easily through on-demand autoscaling and load-balancing in the cloud or datacenter environments~\cite{qiu2020firm,rzadca2020autopilot,singh2019research,novak2020auto,novak2019cloud}.

Second, sensitivity characterization in CoCo is workloads or application inputs dependent, and can not fully observe all system states.
For instance, sensitivity can vary with non-modeled system states such as PCIe bandwidth contention or disk I/O contention.
Essentially, sensitivity should be a distribution among the factors of workloads and system states, in the presence of hidden resources.
To address that, reinforcement learning based approaches~\cite{qiu2020firm,banerjee2020inductive} under partially observable environments provide possible directions without using painstakingly tuned heuristics.

% 1. sensitivity does not change by other non-modeled systems resources; i.e., slowdown caused by CAT/MBA limiting is not affected by other systems resource constraints
% 2. sensitivity characterization is workload dependent (or input to the application)
% 3. sensitivity should be a distribution "f(workload characteristics, system states)", but currently CoCo only works for a system state snapshot; to work for all cases, CoCo needs to have all those states profiled (or use a dynamic RL-approach)

%% file: 004-eval.tex
\section{Preliminary Results}
\label{sec:eval}

Our evaluation is conducted on an application-hosting server with an AMD EPYC 7302P processor with 16 cores and 126 GB of memory, which supports 4 CLOSs with 20-way CAT partitioning and 0-100\% range of MBA throttling.
We run co-located applications on this server.
We use a client machine to create workloads by sending client requests to the application server. The client server has an Intel(R) Core(TM) i7-7700 processor with 8 CPUs and 16 GB memory.
% conducted on two physical servers, one application server serving  and one client server used solely for sending client requests to the application server.
% To study the effect of NUMA nodes, we also run experiments on an application server with an Intel(R) Xeon(R) CPU E5-2695 v4 processor, which only supports Intel CAT but has two NUMA nodes.

% The servers are running Linux version 5.4.0-54, 4.15.0-123, and 4.15.0-70 respectively.

% \begin{figure*}[!ht]
%     \centering
%     \subfigure[Impact of Intel CAT Throttling]{\includegraphics[scale=0.5]{figs/cat-throttling-impact.pdf}}\quad
%     \subfigure[Impact of Intel MBA Throttling]{\includegraphics[scale=0.5]{figs/mba-throttling-impact.pdf}}
%     \caption{Impact of Intel CAT/MBA on Different Applications}
%     \label{fig:slowdown}
% \end{figure*}

We choose three representative LC applications, Nginx~\cite{nginx}, Memcached~\cite{memcached}, and MongoDB~\cite{mongodb}.
To drive the workloads, we use wrk2~\cite{wrk2}, memtier~\cite{memtier}, and YCSB~\cite{ycsb}.
We run each application in a separate Docker container on the application server with the workload generation benchmarks set up in the client server.
% The Docker version we use is 20.10.0.
No other background jobs are running on both machines.

% Memcached~\cite{memcached} is a high-performance in-memory key-value store, which is commonly used as a distributed memory object caching system.
% It has become a critical tier in cloud services to provide low-latency.
% We use memtier\_benchmark~\cite{memtier} to generate various traffic patterns of workloads, as it provides a robust set of customization and reporting capabilities.
% In particular, we choose a sequential access pattern to random data with balanced SET/GET ratio.
% Nginx~\cite{nginx} is a high-performance web server that powers 400 million websites.
% We use it as the font-end to serve $10^4$ static files of size 1.5 KB each.
% We use wrk2~\cite{wrk2} as the open-loop workload generator to generate web page access traffic, where the number of threads, client connections, and request rates are all configurable.
% MongoDB~\cite{mongodb} is the most popular NoSQL database widely used in industry as back-end data storage.
% We use Yahoo! Cloud Serving Benchmark (YCSB)~\cite{ycsb} as the workload generator where read/write mix ratio, request distribution, and record size can be defined flexibly.
% In particular, we use workload-A where the ratio between read-requests and write-requests is 1:1.

% \textbf{Resource Contention Injection.}
% To characterize the performance degradation of each application under shared resource contention, we modified iBench\cite{ibench} to inject noisy background jobs.
% Benchmarks in iBench can create contention on last-level cache (LLC) and memory bandwidth with tunable intensity.

\begin{table}[!ht]
    \centering
    \caption{Impact of CAT/MBA on LC applications in terms of load retainment (normalized to full allocation).}
    \vspace{5pt}
    \label{table:characterization_cat}
    \resizebox{\linewidth}{!}{%
        \begin{tabular}{lrrrr}
        \toprule
        CAT Config. & \textbf{Full} & \textbf{9-bit Mask} & \textbf{6-bit Mask} & \textbf{3-bit Mask} \\ 
        \midrule
        \textbf{Memcached} & 100\% & 88.1 $\pm$ 5\% & 83.8 $\pm$ 2\% & 80 $\pm$ 4\% \\
        \textbf{Nginx} & 100\% & 75 $\pm$ 4\% & 62 $\pm$ 3\% & 33 $\pm$ 3\% \\
        \textbf{MongoDB} & 100\% & 58.3 $\pm$ 5\% & 37.3 $\pm$ 1\% & 26 $\pm$ 2\% \\
        \bottomrule
        \vspace{-8pt}
        \end{tabular}%
    }
    \resizebox{\linewidth}{!}{%
        \begin{tabular}{lrrrrr}
        \toprule
        MBA Config. & \textbf{Full} & \textbf{80\%} & \textbf{60\%} & \textbf{40\%} & \textbf{20\%} \\ 
        \midrule
        \textbf{Memcached} & 100\% & 91.4 $\pm$ 6\% & 87.2 $\pm$ 4\% & 82 $\pm$ 5\% & 78.4 $\pm$ 5\%\\
        \textbf{Nginx} & 100\% & 93 $\pm$ 4\% & 90.1 $\pm$ 3\% & 87.3 $\pm$ 2\% & 81.1 $\pm$ 4\% \\
        \textbf{MongoDB} & 100\% & 82.5 $\pm$ 5\% & 74 $\pm$ 3\% & 69.9 $\pm$ 4\% & 64.2 $\pm$ 4\%\\
        \bottomrule
        \end{tabular}%
    }
\end{table}
% \begin{table}
%     \centering
%     % \vspace{-10pt}
%     \caption{Impact of Intel MBA on different applications.}
%     % \vspace{-5pt}
%     \label{table:characterization_mba}
%     \resizebox{\linewidth}{!}{%
%         \begin{tabular}{lrrrrr}
%         \toprule
%         Load Retainment & \textbf{Full} & \textbf{80\%} & \textbf{60\%} & \textbf{40\%} & \textbf{20\%} \\ 
%         \midrule
%         \textbf{Memcached} & 100\% & 91.4 $\pm$ 6\% & 87.2 $\pm$ 4\% & 82 $\pm$ 5\% & 78.4 $\pm$ 5\%\\
%         \textbf{Nginx} & 100\% & 93 $\pm$ 4\% & 90.1 $\pm$ 3\% & 87.3 $\pm$ 2\% & 81.1 $\pm$ 4\% \\
%         \textbf{MongoDB} & 100\% & 82.5 $\pm$ 5\% & 74 $\pm$ 3\% & 69.9 $\pm$ 4\% & 64.2 $\pm$ 4\%\\
%         \bottomrule
%         \end{tabular}%
%     }
% \end{table}

\subsection{Sensitivity of Co-located Workloads}
\label{sec:eval-characterization}

% talk about impact of Intel CAT and MBA throttling
% \textbf{Impact of Intel CAT and MBA Throttling.}
As discussed in \S\ref{sec:overview-profiling},
different applications have different sensitivities to resource contention.
% different types of shared resource contention, and (b) Different applications have different sensitivities to the same type of shared resource contention.
To cater to different sensitivities of different workloads, we need to understand how Intel CAT and MBA throttling will affect the performance.
Based on the characterization, we can then compute the weights used in the MQ-WRR dynamic resource partitioning algorithm (recall \S\ref{sec:overview-algorithm}).

Table \ref{table:characterization_cat} show the results.
The max load is measured without associating any CAT mask (i.e., full-mask) or without adding any MBA throttling rate (i.e., 100\% bandwidth) to the workload.
We gradually increase the load of the client requests to a point at which the pre-defined SLO is violated within a 95\% confidence interval.
Then we choose 9-bit, 6-bit, and 3-bit masks as three examples to calculate the slowdown values for LLC, and choose 80\%, 60\%, 40\%, and 20\% throttling rates to calculate the slowdown values for MBA.

To do so, we pin each co-located workload to a set of cores (we assign no fewer cores than threads to avoid the CPU being the bottleneck) by changing the \texttt{cpuset.cpus} of the \texttt{cgroup} configuration for that container, and then associate the corresponding CAT and MBA CLOSs to the cores.
After configuration, we gradually increase the load of client requests to a point at which the SLO of the application is violated and mark it as $L$.
The ratio of that load $L$ to the max load (i.e., with no throttling) is then the load retainment value shown in the table.

\begin{figure}[!t]
    \centering
    \vspace{-1pt}
    \includegraphics[scale=0.525]{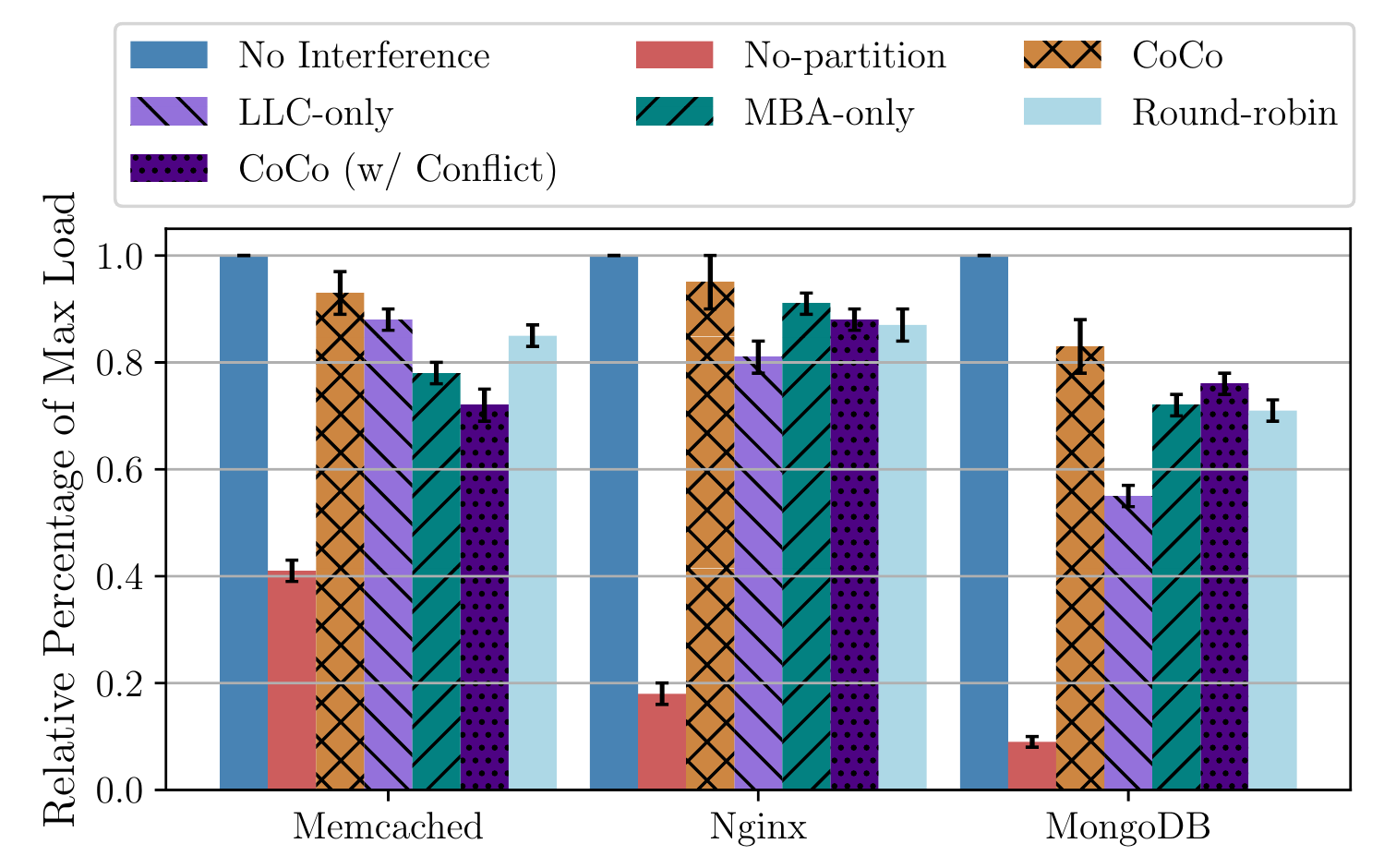}
    \vspace{-15pt}
    \caption{Affordable client load without SLO violations (normalized to no-interference, which is 100\%).}
    % \vspace{-10pt}
    \label{fig:e2e}
\end{figure}

\subsection{Effectiveness of CoCo}
\label{sec:eval-e2e}

% talk about e2e evaluation - load retainment improvement
We show the effectiveness of CoCo by comparing it with (i) no partitioning and (ii) standard round-robin scheduling~\cite{tavakkol2018flin}.
%, each of which is deployed in a separated Docker container.
Since the physical server on which the co-located workloads sit is able to configure a max of 4 CLOSs and in total 20 LLC ways (20-bit mask), we divide the LLC ways into 4 CLOSs, each of which is allocated 2, 3, 6, 9 bits of LLC masks and 10\%, 10\%, 30\%, 50\% of memory bandwidth respectively.
The co-located workloads are two Memcached instances, two Nginx instances, and two MongoDB instances.
The reason why we did not compare with static partitioning is that there is no way for six workloads to statically share four CLOSs.
To show the benefit of coordinating CAT and MBA together, we added the evaluation of CoCo with only CAT, only MBA, and conflicting CAT/MBA.

Figure \ref{fig:e2e} shows the results comparatively.
CoCo improves the load retainment compared with no-partitioning by 2.2--9.1$\times$.
We summarize several key results: (a) CoCo outperforms CAT-only or MBA-only dynamic resource partitioning by coordinating CAT and MBA together.
(b) CoCo schedules the partition in CAT to be positively correlated with the partition in MBA (never reduces the LLC partition when increasing the memory partition) and thus outperforms CoCo with conflicts.
(c) CoCo outperforms standard round-robin scheduling by considering the slowdown value of each workload. Round-robin wastes resource partitions on unnecessary workloads and ignores workloads that are more sensitive.
% talk about the overhead of CoCo brought by context switch and time sharing
While achieving the improvement on load retainment, CoCo brings overhead to the end-to-end latency of 2.4--6.1\% due to the partition context switch and cache flush during CLOS migration.
% We will further discuss this in Section \ref{sec:discussion}.

%% file: 005-discussion.tex
\section{Related Work}
% \textbf{QoS-Aware Resource Management}
Using resource partitioning to alleviate contention of shared resources between co-located applications is well studied~\cite{heracles,parties,clite,copart,qiu2020firm}.
For example,
Heracles~\cite{heracles} presents a feedback-based resource controller that enables an LC task to be co-located with any number of best-effort (BE) tasks without breaking the SLO of the LC task.
However, it can support at most one LC task.
PARTIES~\cite{parties} and CLITE~\cite{clite} both provide a QoS-aware dynamic resource manager which allows a number of LC applications and BE tasks to safely share one commodity server without violating SLOs.
% Although they do not require offline profiling or any prior knowledge of applications, the controller decision sometimes can be counterproductive, which results in low resource utilization and QoS violation.
% Without application profiling, it takes long time for the above-mentioned controller to converge (more than 30s)~\cite{fried2020caladan}.
CoPart~\cite{copart} dynamically analyzes the characteristics of the co-located applications and
partitions the LLC and memory bandwidth in a coordinated manner to maximize the overall fairness of the
applications.
The number of LC applications that the above-mentioned approaches can support is also limited by the hardware features (i.e., the number of CLOSs in Intel CAT/MBA).

\section{Concluding Remarks}
\label{sec:conclusion}
% talk about the limitation of hardware-based resource partition
% increasing the number of CLOS still not solve the problem -- e.g., which is limited by the number of ways for way-based cache partition

CoCo is the first step in the direction towards guaranteeing SLOs for more LC applications beyond the hardware allocation capability by treating partitions as scarce resources and designing sharing policies.

The root of the problem is the scarcity of CLOSs supported by existing hardware QoS support.
% ---four CLOSs obviously fall short in the face of the scale of CPU cores and the size of memory.
It is perhaps time to rethink hardware QoS support.
According to our understanding, the main challenge for way partitioning to support more applications comes from the difficulty of scaling cache associativity to the number of cores, as physical constraints result in increased latency and energy consumption; therefore, simply increasing the number of CLOSs cannot solve the problem because the cache associativity would become the bottleneck.
On the other hand, it is hard to partition below the way level~\cite{xiang2018dcaps,sanchez2010zcache}.

In summary, we believe that the discrepancy between the hardware QoS support and the scale of CPU cores and memory will continue to exist if not become worse.
Software-based solutions like CoCo that exploit the CLOS shareability will play an important role in providing SLOs and improving hardware resource utilization.